\documentclass[twocolumn]{aastex6}

\begin{document}

\title{The First Scattered Light Image of the Debris Disk around the Sco-Cen target HD~129590}

\author{Elisabeth Matthews\altaffilmark{1}}
\author{Sasha Hinkley\altaffilmark{1}}
\author{Arthur Vigan\altaffilmark{2}}
\author{Grant Kennedy\altaffilmark{3}}
\author{Aaron Rizzuto\altaffilmark{4}}
\author{Karl Stapelfeldt\altaffilmark{5}}
\author{Dimitri Mawet\altaffilmark{6,5}}
\author{Mark Booth\altaffilmark{7}}
\author{Christine Chen\altaffilmark{8}}
\author{Hannah Jang-Condell\altaffilmark{9}}

\altaffiltext{1}{University of Exeter, Astrophysics Group, Physics Building, Stocker Road, Exeter, EX4 4QL, UK}
\altaffiltext{2}{Aix Marseille Université, CNRS, LAM (Laboratoire d’Astrophysique de Marseille) UMR 7326, 13388, Marseille, France}
\altaffiltext{3}{Institute of Astronomy, University of Cambridge, Madingley Road, Cambridge CB3 0HA, UK}
\altaffiltext{4}{Department of Astronomy, The University of Texas at Austin, Austin, TX 78712, USA}
\altaffiltext{5}{Jet Propulsion Laboratory, California Institute of Technology, 4800 Oak Grove Drive, Pasadena, CA 91109, USA}
\altaffiltext{6}{California Institute of Technology, Division of Physics, Mathematics and Astronomy, 1200 E. California Blvd, Pasadena, CA 91125, USA}
\altaffiltext{7}{Astrophysikalisches Institut und Universitatssternwarte, Friedrich-Schiller-Universit{\"a}t Jena, Schillerg{\"a}{\ss}chen 2-3, D-07745 Jena, Germany}
\altaffiltext{8}{Space Telescope Science Institute, 3700 San Martin Drive, Baltimore, MD 21218, USA}
\altaffiltext{9}{Department of Physics \& Astronomy, University of Wyoming, Laramie, WY 82071, USA}

\begin{abstract}

We present the first scattered light image of the debris disk around HD~129590, a $\sim$1.3~M$_\odot$~G1V member of the Scorpius Centaurus association with age $\sim$10-16~Myr. The debris disk is imaged with the high contrast imaging instrument SPHERE at the Very Large Telescope, and is revealed by both the IRDIS and IFS subsytems, operating in the H and YJ bands respectively. The disk has a high infrared luminosity of $L_{\textrm{IR}}/L_{\textrm{star}}$$\sim$5\,$\times$\,10$^{-3}$, and has been resolved in other studies using ALMA. We detect a nearly edge on ring, with evidence of an inner clearing. We fit the debris disk using a model characterized by a single bright ring, with radius $\sim$60-70~AU, in broad agreement with previous analysis of the target SED. The disk is vertically thin, and has an inclination angle of $\sim$75$^\circ$. Along with other previously imaged edge-on disks in the Sco-Cen association such as HD~110058, HD~115600, and HD~111520, this disk image will allow of the structure and morphology of very young debris disks, shortly after the epoch of planet formation has ceased.

\end{abstract}


\section{Introduction} \label{sec:intro}

Observing the youngest stellar systems, shortly after planet formation has ceased, provides a glimpse of nascent circumstellar environments. However, only a small handful of stars ($\lesssim$3-5) with ages $\lesssim$10-20\,Myr, just after dissipation of the gaseous primordial disc, are found within 100\,pc \citep{sacco2014}. The Scorpius Centaurus Association \citep[][hereafter Sco-Cen]{dezeeuw1999} is the nearest OB2 association, with a mean distance of 140\,pc, making it perhaps the most promising collection of young stars which can be observed shortly after the period of active planet formation. This region, containing stars with ages $\sim$~10-16\,Myr \citep{pecautmamajek2016} allows the best constraints to be placed on the orbital zones of planet formation, and on the early thermal histories of young planets \citep{ireland2011,janson2013,lafreniere2011,hinkley2015b}.

Furthermore, such young systems often possess bright circumstellar debris disks, belts of planetesimals analogous to the Edgeworth-Kuiper belt in our own solar system. The presence of dust in these systems suggests that planetesimals are responsible for the dust generation \citep{wyatt2008}. The dust that is generated through planetesimal collisions in these debris disks is inherently transient, being either blown out by stellar winds, or spiralling towards the host star via Poynting-Robertson drag.  The persistence of dust in these systems implies that it is constantly being regenerated.  Although populations of planetesimals may stir themselves in some cases \citep[e.g.][]{kennedywyatt2010}, this dust regeneration may be enhanced by perturbations from massive planets, dynamically exciting the planetesimals onto eccentric orbits, thus causing them to collide.  As well as revealing the presence of massive planetary perturbers, the location of these large quantities of dust may hint at the location of giant planets in the system: gaps between dust belts may highlight where planets lie \citep{surieke2014}, and sharp edges to debris rings can constrain the masses of planets shepherding these edges \citep[e.g.][]{quillen2006,chiang2008,mustillwyatt2012}. 

\begin{figure}
\label{sed}
\plotone{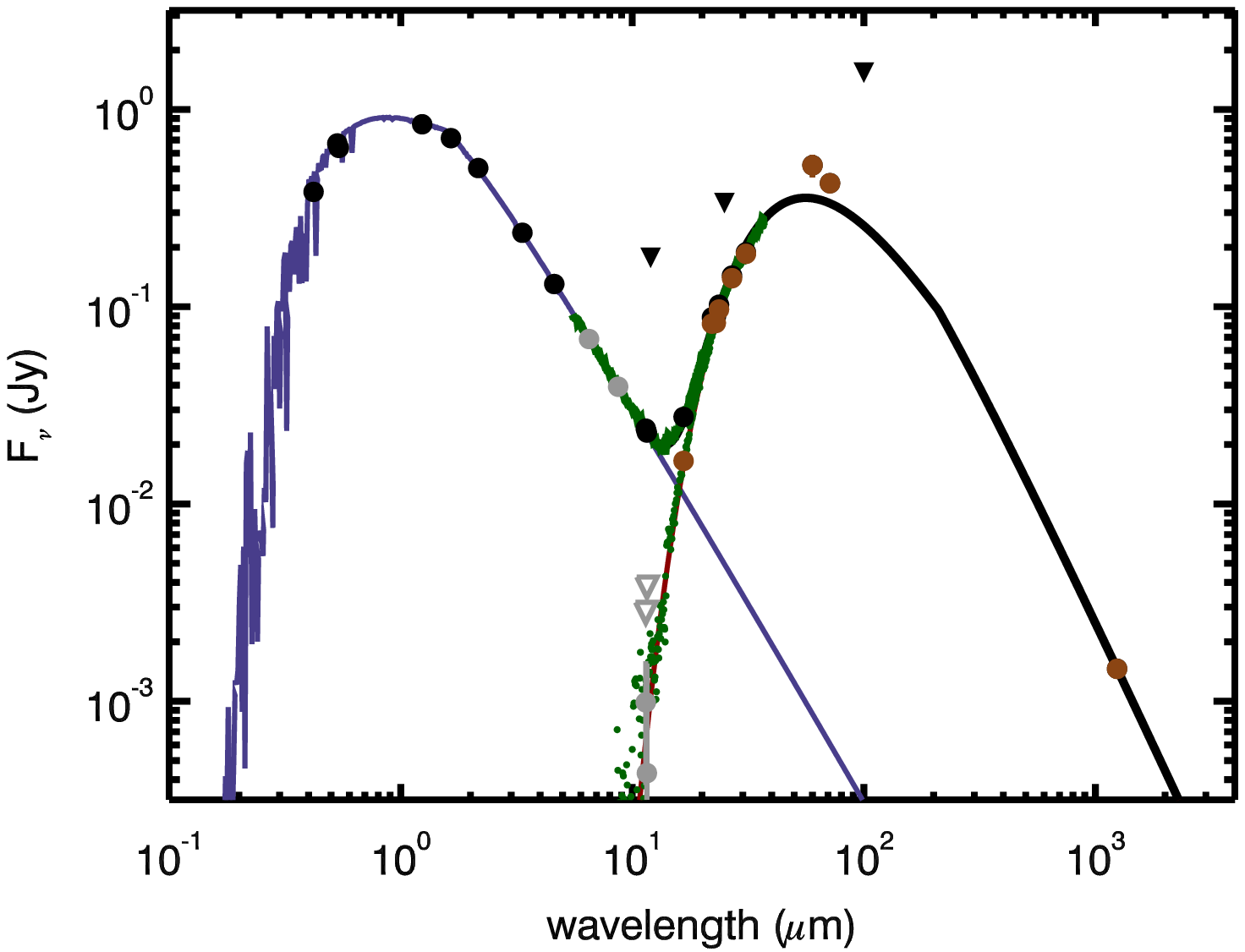}
\caption{Spectrum of HD~129590. Black dots show photometry from \emph{Hipparcos}, 2MASS, WISE, IRAS, \emph{Spitzer} and ALMA \citep{hog2000, cutri2003, wright2010, helou1988, chen2014, liemansifry2016}. Triangles show IRAS upper limits, and small green dots show the \emph{Spitzer} IRS spectrum. The lines show a 5850K PHOENIX stellar model \citep[blue,][]{brott2005} and a 91K modified blackbody (black). Brown and green dots show star-subtracted measurements (which cover the black dots in some cases). Grey open triangles indicate where star-subtracted values are consistent with zero.}
\end{figure}

One Sco-Cen system with a known circumstellar debris disk is HD~129590 \citep[HIP~72070, T$_{\textrm{eff}}$~=~5945~K, 1.3~M$_\odot$ 2.8~L$_\odot$, ][]{chen2011}. While not originally catalogued in the \citet{dezeeuw1999} or \citet{rizzuto2011} catalogs of Sco-Cen stars, HD~129590 has been listed as a G1V member of the Sco-Cen subgroup Upper Centaurus Lupus \citep{hoogerwerf2000,chen2011}. HD~129590 has an estimated distance of 141$\pm$7~pc \citep{astraatmadja2016,gaia_dr1a,gaia_dr1b} and a high infrared luminosity $L_{\textrm{IR}}/L_{\textrm{star}}$$\sim$5\,$\times$\,10$^{-3}$, twice the value observed for $\beta$~Pictoris \citep{jangcondell2015,mittal2015}, as can be seen in Figure \ref{sed}. Although \citet{chen2014} fit the SED as two distinct belts, \citet{jangcondell2015} suggest the system likely contains a single belt of dust. This is in agreement with the predictions of \citet{ballering2013} (see \S \ref{sed-fit} for further details).

The high fractional luminosity of HD~129590 makes it an extremely promising target for scattered light imaging of circumstellar material \citep{currie2014,draper2016}. ALMA data have previously been obtained for this object \citep{liemansifry2016}, with the disk being marginally resolved along the major axis. \citet{liemansifry2016} constrain the inclination angle to $>$\,50$^{\circ}$ with a best fit value of 70$^{\circ}$, and the position angle to -59$^{+17}_{-12}\,^{\circ}$. The ALMA data finds a best-fit grain size of $3.2^{+0.6}_{-0.5}$~\micron, and at this size find the outer edge to be $110^{+50}_{-30}$~AU and the inner edge to be $<$\,40~AU. No gas was detected in that work, and so we do not expect gas to have a strong influence on the dust dynamics in this disk.

In this letter, we present the first scattered light image of the debris disk around HD~129590, using the SPHERE instrument on the VLT. The observations and data post-processing are presented in \S \ref{sec:obs} and \S \ref{sec:dr} respectively. In section \S \ref{sec:model} we describe our modelling of the disk, where we use an optically thin disk model to conclude that the disk has a radius of $\sim$~60-70~AU. Finally, we conclude in \S \ref{sec:conclusion}.

\section{Observations} \label{sec:obs}

HD~129590 was observed on 2016~May~4 with the SPHERE instrument at the VLT \citep{beuzit2008}, as part of a larger planet-finding survey within Sco-Cen. The data were taken in the IRDIFS mode, whereby light is split through a dichroic beamsplitter, and passed simultaneously to both the differential imager and spectrograph instrument \citep[IRDIS;][]{dohlen2008}, and the integral field spectrometer \citep[IFS;][]{claudi2008}. A total of 2560s on-sky integration was collected by each instrument, with the \texttt{N\_ALC\_YJH\_S} coronagraph in place. The sequence consisted of 80$\times$32s individual exposures for the IRDIS data and 40$\times$64s exposures for the IFS data. We used IRDIS in dual-band imaging \citep[DBI;][]{vigan2010} mode with the H23 filter pair, at wavelengths $\lambda$\,=\,1588.8\,nm, $\Delta \lambda$\,=\,53.1\,nm and $\lambda$\,=\,1667.1\,nm, $\Delta \lambda$\,=\,55.6\,nm, while the IFS instrument \citep{zurlo2014,mesa2015} was used in the \texttt{YJ} mode, which spans the range 0.95-1.35\,$\micron$. Platescales of the instruments are 12.25~mas/pix for IRDIS and 7.46~mas/pix for the IFS \citep{maire2016}, and the inner working angle of the coronagraph is 0.15$^{\prime\prime}$.

In addition to these science observations, flux calibration images were collected with the target displaced from the coronagraph. Star center calibration frames (waffle frames) were also collected, by imposing a sinusoidal pattern on the deformable mirror. This creates four starspot images, with equal displacements from the central star, in each corner of the frame. Together, these allow the star position to be accurately measured to $\sim$0.1~pixels (1.2~mas) \citep{vigan2016} behind the occulting mask. The observations were carried out in pupil-stabilized mode to allow angular differential imaging analysis \citep[ADI;][]{marois2006}. The entire sequence of observations, including acquisition and calibration, lasted 59 minutes spanning an airmass range of 1.037 to 1.042.  The primary science frames covered a total field rotation of 36 degrees, and included the meridian crossing of the target.

\section{Data Post-Processing} \label{sec:dr}

Our data post-processing was carried out following the process described in \citet{vigan2015sirius} (here on: V15). We used both the ESO data reduction and handling pipeline \citep[DRH,][]{pavlov2008} and the publicly available code described in V15, as well as some custom routines.

\begin{figure*}
\label{irdis-ifs}
\centering
\includegraphics[width=0.95\textwidth, clip=true]{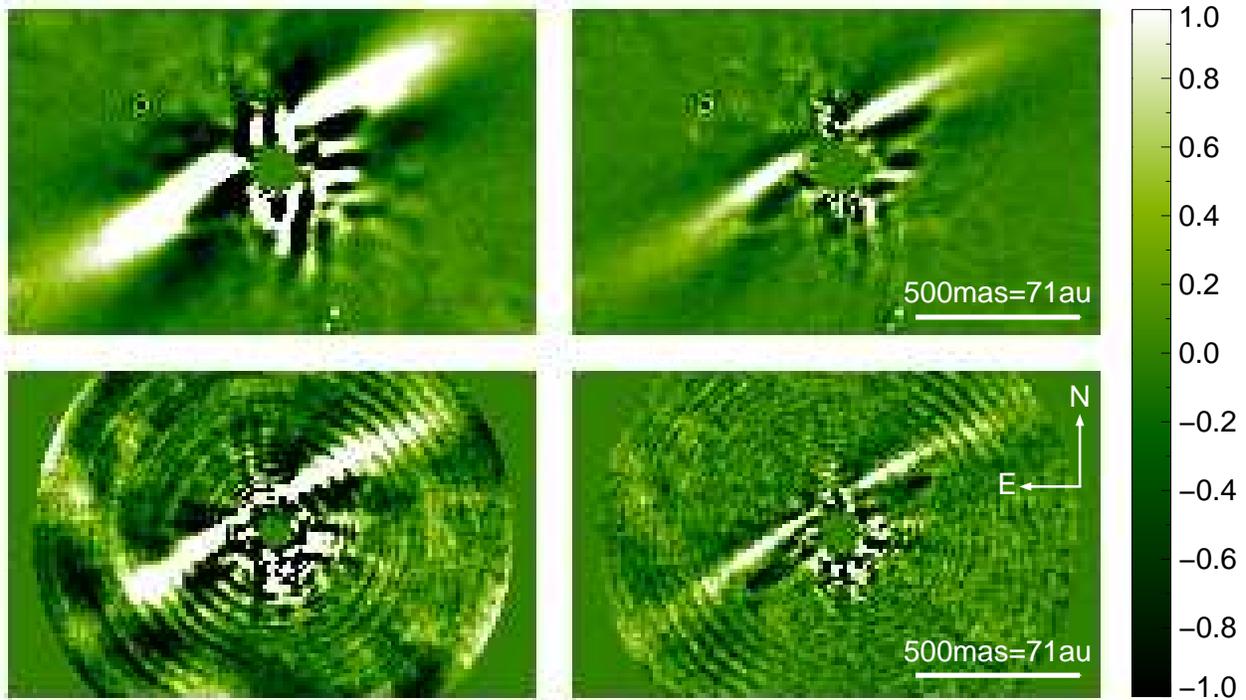}
\caption{SPHERE detections of a scattered light debris disk around the target HD~129590. Data from the IRDIS (1.6\,$\micron$) and IFS (0.95-1.35$\micron$) subsystems are shown on the top and bottom rows respectively. In each case, the image is a co-add of the entire wavelength range of the subsystem. The left hand images have 6 principal components subtracted, while on the right hand side a more aggressive reduction is presented, where 20 principal components have been removed. In both cases, the reduction is full-frame treatment, where the entire field of view is considered simultaneously. The ringed structure observed in the IFS images is an artifact of the fast Fourier transform process used to rescale the various wavelength observations.}
\end{figure*}

\subsection{IFS}

For the IFS data, basic calibrations were first created using the DRH: dark fields, master flat-fields, IFS spectra positions, initial wavelength calibrations and an IFU flat were all generated. We then used a custom routine to calculate an accurate parallactic angle and time for each image, and to normalise the data based on its exposure time. Bad pixel and cross-talk corrections were applied (for details see V15). The DRH was then used to interpolate these frames both spectrally and spatially. A sigma-clipping routine was applied to remove remaining bad pixels which deviated from their neighbours by more than 3.3$\sigma$. Finally, the wavelengths for each image were recalibrated, due to small systematic errors in the DRH pipeline, as described in V15. This process results in a set of calibrated images in the x-y plane, at 39 wavelengths spanning 0.95-1.35~$\micron$ and at 40 distinct timesteps. No frame selection was performed, although each individual frame was visually inspected to confirm there were no data issues.

Speckle subtraction was performed using a custom PCA code \citep[e.g.][]{soummer2012,amaraquanz2012}. This code simultaneously uses the spectral and temporal (parallactic angle) diversity of speckles to remove starlight scattered within the image plane by the telescope optics, but not genuine astrophysical sources. A disk feature was revealed, as shown in Figure \ref{irdis-ifs}. We tested reductions with between 2 and 100 principal components subtracted, and the feature is robust to the number of principal components removed.

\subsection{IRDIS}

Initial pre-processing of the IRDIS data was performed using the ESO SPHERE pipeline. Master dark and flat frames were created, and a waffle frame was used to calibrate the position of the star centers. Each frame was then independently reduced by applying the master dark and flat frames, and realigned using the star center calibrations and the dither positions so that the central star position was consistent between images.

The 160 individual images (80 timesteps, 2 wavelengths) were then input into a PCA algorithm \citep[e.g.][]{soummer2012,amaraquanz2012} to remove stellar speckle noise. For this process we used our own modification of the V15 IFS code, which was reconfigured to accept IRDIS data. A clear disk feature was observed, as demonstrated in Figure \ref{irdis-ifs}. This disk feature closely matches that observed in the IFS data. As with the IFS data, we tested a range of reductions with between 2 and 100 principal components removed and found the disk feature to be robust. 

\section{Disc Structure \& Modelling} \label{sec:model}

\subsection{SED fitting}
\label{sed-fit}

The infrared excess of HD~129590 is well-studied. \citet{ballering2013} fit a single cold dust component at a temperature of 89~K, while \citet{jangcondell2015} find a best fit with grain temperature 93.7$\pm$0.1±K. \citet{chen2014}, however, fit two separate components at 94~K and 72~K. The evidence for the second, cold belt comes from a single photometric point at 70~$\micron$ (see SED in Figure \ref{sed}), which might be equally well explained by a dust model that is more complex than a simple blackbody. We conclude that the SED is best and most simply described by a single blackbody, with a best-fit temperature of 91~K (and radius 16~AU assuming blackbody absorption/emission). Small dust grains emit poorly at wavelengths significantly longer than their physical size, which we parametrise by “modifying” our blackbody fit by a factor $(210 \mu {\rm m}/\lambda)^\beta$ at wavelengths longer than 210$\mu$m \citep{wyatt2008}. This extra parameter is required by the ALMA observations, and $\beta=0.5$ yields the best fit.

\subsection{Spatial Constraints}

The debris disk is highly symmetric in both the IFS and the IRDIS data. In the IRDIS data, there is a clear dark hole within 0.32$^{\prime\prime}$ of the star. The brightest emission extends to 0.67$^{\prime\prime}$, but there is evidence of extended emission as far as 1.03$^{\prime\prime}$ from the star, in line with the debris disk. Both lobes of the disk can be clearly seen. The disk signal in the IFS data is fainter, with only the front lobe visible. This has a spatial extent of 0.57$^{\prime\prime}$. As in the IRDIS data, there is extended, faint emission to the edge of the the field of view (0.87$^{\prime\prime}$).

\subsection{Disk Modelling}

We then use injection modelling to characterise the disk more rigorously. We choose to take a Bayesian MCMC fitting approach, as performed in \citet{wahhaj2014}.

Synthetic disk images are created using the GRaTeR radiative transfer code \citep{augereau1999}. We use an optically thin disk model with density defined as:

\begin{equation}
    \rho = \frac{\rho_{\circ} \exp  \left( \left[\frac{-|z|}{\xi_\circ}\left(\frac{r}{r_{\circ}}\right)^{-\beta}\right]^\gamma \right) }
    {\sqrt{\left(\frac{r}{r_{\circ}}\right)^{-2\alpha_\mathrm{in}}+
        \left(\frac{r}{r_{\circ}}\right)^{-2\alpha_\mathrm{out}}}}
\end{equation}

where $r$ is the radial distance from the disk center, $z$ is the vertical distance from the disk midplane, and $\rho_{\circ}$, $\alpha_\mathrm{in}$, $\alpha_\mathrm{out}$, $\xi_\circ$, $\beta$ and $\gamma$ are free parameters. The exponential term defines the disk vertical profile, while the denominator expresses a radial profile which rises as $\alpha_\mathrm{in}$, peaks at $r_{\circ}$ and then falls as $\alpha_\mathrm{out}$.

At each point in the disc, the scattered light contribution is calculated as:

\begin{equation}
F \propto \frac{\rho \times \textrm{p}(\theta)}{d^2}
\end{equation}

where $\theta$ is the scattering angle, namely, the angle through which light from the star is scattered so that it reaches the Earth. $d$ is the distance from the star to the grid point. The measured flux is influenced by several factors, such as the stellar luminosity, stellar distance and the telescope gain. We fold these into a single overall normalization by modifying the $\rho_{\circ}$ parameter and denote this updated parameter as $\rho_{\circ}^{\prime}$.

The phase function, $\textrm{p}(\theta)$, is the standard Henyey-Greenstein scattering function:

\begin{equation}
    \textrm{p}\left(\theta\right) = \frac{1}{4\pi} \frac{1-g^2}{\left[1-2g\cos{\theta}+g^2\right]^{\frac{3}{2}}}
\end{equation}

Since we assume an optically thin disc, scattered light contributions are added along each line of sight to create a synthetic disk image.

These synthetic images were convolved with a Gaussian to mimic the effects of the point spread function. Each model in turn was then rotated and subtracted from each frame of the raw IRDIS data, and the PCA sequence was repeated with six modes subtracted to generate a residual image. By injecting negative disk images, we accurately take into account the throughput of the PCA post-processing, which varies with separation from the star. This process is akin to a forward modelling procedure. In the interests of computational efficiency, we use only a subset of the data for our modelling: images are trimmed to the central 240$\times$240 pixels ($\sim$3$^{\prime\prime}$), and only every fourth individual exposure is included.

The goodness of each model is assessed using the normal $\chi^2$ statistic: each pixel in the residual image is divided by its local sigma value and squared. Values for $\sigma$ are calculated as in \citet{wahhaj2014}: we first convolve the PCA processed image of the disk with a Gaussian with FWHM of 2 pixels. This is then subtracted from the image to remove extended spatial components, to leave an image containing only noise information. The rectangular test region is divided into annuli, and in each annulus the standard deviation of the noise image is found, so as to capture the variance of noise with distance from the star. As noted in \citet{wahhaj2014}, some disc signal still contributes to the standard deviation. This is inevitable, and will lead to conservative error calculations on our parameters. 

We initially use a downhill minimization routine to find a best fit. We then use these best fit parameters to initiate a Metropolis Hastings MCMC, using the \texttt{emcee.py} package \citep{foremanmackey2013}. The MCMC chain generates a sampling of our parameters space, with probabilities assigned as $\exp \left(\frac{-\chi^2}{2} \right)$. We use uniform priors in this work. 510,000 random samplings are generated for each fit, and the first 10,000 are discarded to ensure that the results are independent of our starting position. The best fit and error values are calculated from the remaining 500,000 samples. For each parameter, the best fit given by the median value of the marginalized distribution and the 1-$\sigma$ uncertainties chosen to enclose 34\% of the samples on either side of the median. In addition, we calculate the best fit parameters for each third of the random samplings, and find them to be consistent. This confirms that the initial parameters do not affect the results, and that they are a genuine sample of the probability distribution.

\begin{figure*}
\centering
\includegraphics[width=0.95\textwidth, trim=0.5cm 0.2cm 0.5cm 10.3cm, clip=true]{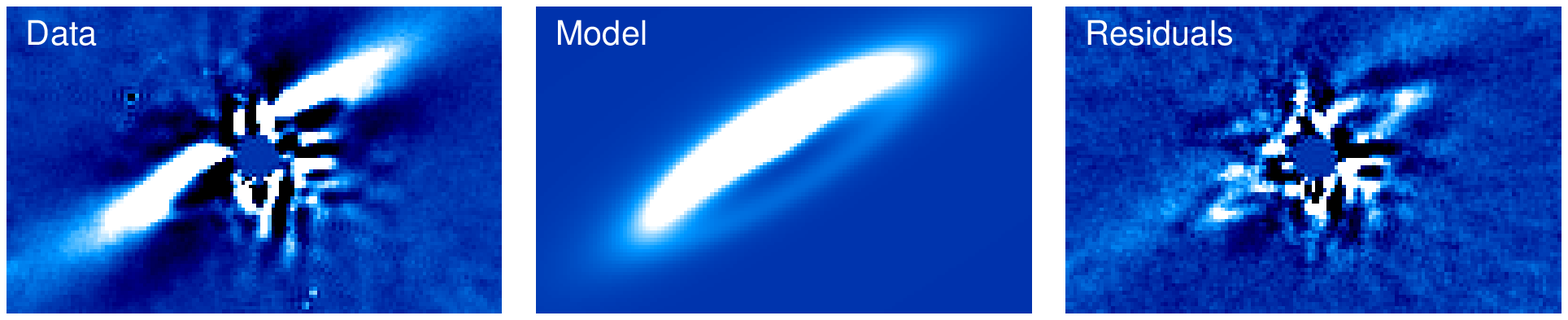}
\includegraphics[width=0.95\textwidth, trim=0.5cm 0.2cm 0.5cm 10.3cm, clip=true]{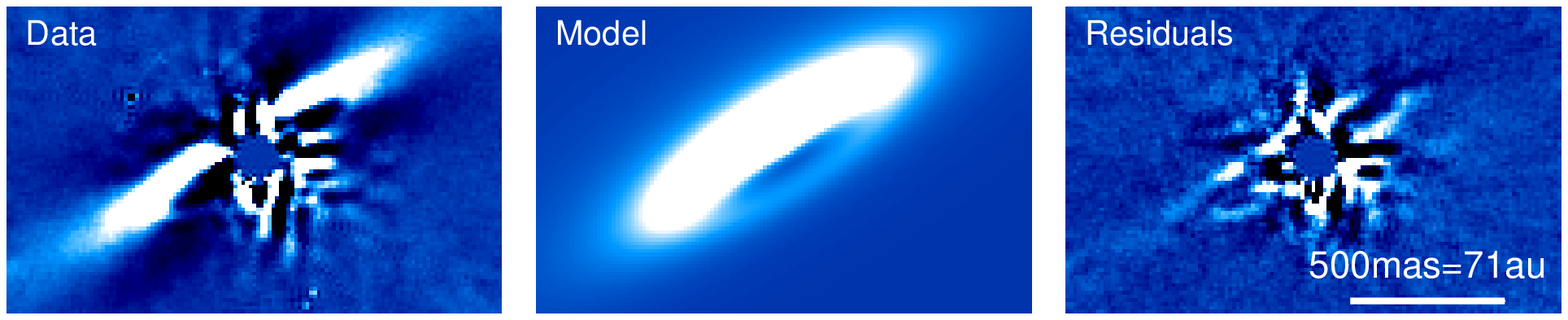}
\caption{\textit{{Top:}} The initial debris disk fit, where the Henyey-Greenstein parameter $g$, the disk radius, and the position and inclination angles of the disk have all been allowed to vary. A residual halo of light is observed. \textit{Bottom:} Here the parameter $\alpha_\mathrm{out}$ has also been allowed to vary. As such, the residual halo is better modelled. Fit parameters for both models are given in Table \ref{tab:fit}. In both cases, IRDIS data is shown with 6 PCA modes subtracted. \vspace{0.5cm}}
\label{fig_fit}
\end{figure*}

We initially fit a single ring of dust. For this model, we have five free parameters: the overall scaling of the model $\rho_{\circ}^{\prime}$, the forward scattering parameter $g$, the radius r$_{\circ}$, the inclination angle and the position angle of the disk. We fix the radial profile to $\alpha_\mathrm{in}=3.5$ and $\alpha_\mathrm{out}=-3.5$ and fix the stellar distance to 141~pc. The parameters defining the vertical profile, namely $\xi_\circ$, $\beta$ and $\gamma$, are fixed to values of 1, 0 and 2 respectively. 
 
This fit is presented in the top row of Figure \ref{fig_fit} with parameters shown in Table \ref{tab:fit}. The residual image for this model shows a clear halo outside the ring, and as such we fit the data again, but this time we also fit the values of $\alpha_\mathrm{in}$ and $\alpha_\mathrm{out}$. The best fit in this case is a smaller ring, with softer power law edges - most notably on the outside, where $\alpha_\mathrm{out}=-1.313^{+0.011}_{-0.012}$. This is a surprisingly low value: \citet{thebaultwu2008} find typical cases of either ``smooth edges" with $\alpha_\mathrm{out}$ -3.5, or ``sharp edges" with $\alpha_\mathrm{out}$ as steep as -8. This second fit is shown in the lower panel of Figure \ref{fig_fit}, with parameters given in Table \ref{tab:fit}. For both of these models, the Henyey-Greenstein parameter takes a relatively high value (0.52 and 0.43, respectively). Although it has been shown for cases with a wide range of viewing angles that a single component Henyey-Greenstein parameter gives a poor fit \citep{stark2014,hedmanstark2015}, the geometry of HD~129590 means that this parameter is poorly constrained: the faint edge is severely affected by speckle noise, meaning backscattering is hard to constrain. 

There is some residual structure that we are unable to model with GRaTeR. In particular, we find no notable improvement in the fit when we vary the disk eccentricity or offset from the star, and the data do not place meaningful constrains on the vertical fit profile. Our position angle is in very good agreement with that found by \citet{liemansifry2016} with ALMA observations, and our inclination angle is within the ALMA constraints.

\begin{deluxetable}{lllllll}
\tabletypesize{\scriptsize}
\tablewidth{8cm}
\tablecaption{Fit parameters for the two disk models in Figure \ref{fig_fit}.
\label{tab:fit}}
\tablehead{Parameter & & Fit 1 & & Fit 2 & & ALMA*} 
\startdata
$\rho_{\circ}^{\prime}$ & & $1.106\pm0.006$ & & $1.440\pm0.008$ & & \\
$g$ & & $0.522\pm0.002$ & & $0.4272\pm0.0012$ & & \\
r$_{\circ}$[AU] & & $73.3\pm0.2$ & & $59.3\pm0.2$ & & \\
i$_\textrm{tilt}$[$^{\circ}$] & & $76.87\pm0.05$ & & $74.56\pm0.05$ & & $>$\,50\\
PA[$^{\circ}$] & & $121.58\pm0.02$ & & $121.80\pm0.02$ & & $121^{+17}_{-12}$\\
$\alpha_\mathrm{out}$ & & -3.5 & & $-1.313^{+0.011}_{-0.012}$ & & \\
$\alpha_\mathrm{in}$ & & 3.5 & & $3.15\pm0.03$ & & \\
$\xi_\circ$ & & 1 & & 1 & & \\
$\beta$ & & 0 & & 0 & & \\
$\gamma$ & & 2 & & 2 & & \\
$\textrm{d}_{star}[pc]$ & & 141 & & 141 & &\\
\enddata
\tablecomments{Values for $\rho_{\circ}^{\prime}$ are relative. PA is measured anti-clockwise of North. *ALMA fit parameters are taken from \citep{liemansifry2016} and converted to our reference system.}
\end{deluxetable}

\section{Discussion}

The latest generation of dedicated high resolution exoplanet imaging platforms such as GPI and SPHERE has already revealed a number of scattered light debris disc images. Sco-Cen has proved to be a particularly fortuitous region for these searches. \citet{currie2015b} detected a dust ring around HD~115600, which was shown to be eccentric. HD~110058 has a wing-tilt asymmetry \citep{kasper2015}, while HD~106906 appears to be misaligned with the wide planetary companion \citep{kalas2015,lagrange2016}. HD~111520 \citet{draper2016} has a dramatic brightness asymmetry, while both HIP~67497 \citep{bonnefoy2017} and HIP~73145 \citep{feldt2017} have been shown to have multiple, separate, rings of debris. Perhaps the only debris discs without complex morphology are HD~114082 \citep{wahhaj2016}  and HD~129590. 

All of these debris discs are presented in \citet{chen2014} as hosting multiple distinct debris belts. Multiple debris belts have only been seen in scattered light around two of the above targets: either there are several very small, close in belts of disc evading detection, or two temperature discs correspond to two belt discs less often than expected. Additionally, all of the targets with debris disc images have very high excess infrared luminosity values, and all except HIP~67497 have been observed with ALMA \citep{liemansifry2016}. 

As discussed in \S 4.1, the debris belt around HD~129590 is predicted to lie at $\sim$16~AU, based on the infrared excess temperature and simple blackbody constraints. Our observations show the ring to peak at  $\sim$4 times this separation, implying an abundance of small dust at higher temperatures than the blackbody temperature. This is to be expected for a luminous star (2.8~L$_\odot$) where the blowout size is a few microns.

\begin{figure}
\label{fig:contours}
\includegraphics[width=0.9\columnwidth, trim=2cm 1cm 4cm 4cm, clip=true]{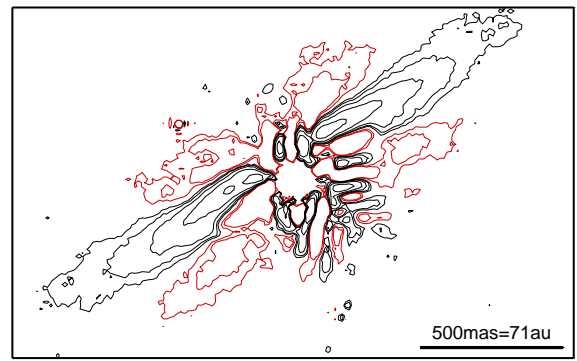}
\caption{The data shown in the top left panel of Figure \ref{irdis-ifs}, but presented as a contour plot. Contours have relative brightness values of 1, 2, 4, 8 and 16, while negative contours are plotted in red, with relative flux -1 and -2. The self-subtraction wings above and below the bright ring edge are clearly visible. No attempt has been made to take the throughput of the PCA routine into account in this plot.}
\end{figure}

We attempt to classify HD~129590 under the categories outlined in \citet{leechiang2016}. The pronounced difference in brightness between the front and back of the disk is reminiscent of the `moth' or `double wing' structures simulated in that work, but we do not observe the bright, extended wings seen in e.g. HD~61005 \citep{hines2007}. A strongly forward-scattering disk could produce only a fainter wing structure, as light will be preferentially scattered away from the viewer. In addition, an ADI based code such as that used here will self-subtract flux in regions at small angular separations to the bright disk edge, and so hide faint wing structures nearby. In Figure \ref{fig:contours} contours of equal brightness are plotted, to demonstrate the full dynamic range of the disk. There is no indication of an extended wing, and indeed the self-subtraction lobes (highlighted in red in the image) show a high degree of symmetry. As such, there is no evidence for moth-like wings and it is not possible to make detailed inferences about the presence, or otherwise, of a planet in this system.

Our modelling work highlighted the presence of extended emission, at large semi-major axis. This could be a dust halo, caused by radiation pressure blowing out small grains. Alternatively, this may be a scattered disk of planetesimals, and as such more closely resemble the wings mentioned above - albeit much more compressed into the plane than \citet{leechiang2016} find. Such a ring of planetesimals might explain our surprisingly low value for $\alpha_\mathrm{out}$. A higher resolution ALMA image could differentiate these two scenarios, while future imaging with a space facility such as the Hubble Space Telescope would allow the lowest surface brightness material (such as a dust halo) to be imaged, as well as defining the outer disk edge far better than ground based AO imaging. Polarimetric differential imaging, meanwhile, would place constraints on the dust grain size and scattering properties, and may even reveal the faint edge of the disc.

\section{Conclusion} \label{sec:conclusion}

HD~129590 is a G1V member of the Sco-Cen association, with an infrared excess twice that observed for $\beta$~Pictoris. This work presents the first scattered light images of the debris disk responsible for this infrared excess. The debris disk is revealed to be a nearly edge-on disk, with evidence for inner clearing. We use the GRaTeR radiative transfer code to model the disk as an optically thin ring, inclined to the line of sight by $\sim$~75$^{\circ}$. Our best fitting model has a characteristic radius of r$_{\circ}$=59.3~AU or r$_{\circ}$=73.3~AU depending on the underlying model, and a forward scattering parameter $g$=0.52 or $g$=0.43. When the power law edges were freed, these were found to take values of 3.15 inside the ring and -1.313 outside the ring. These values imply a strongly forward scattering ring, with a soft outer edge. Even with this model, there is an indication in the final panel of Figure \ref{fig_fit} of some residual structure, implying that there is some morphology more complex than a simple ring present in this disk.

\acknowledgments

Based on observations made with ESO Telescopes at the La Silla Paranal Observatory under programme IDs 095.C-0549 and 097.C-1019. This work has made use of data from the European Space Agency (ESA) mission {\it Gaia} (\url{http://www.cosmos.esa.int/gaia}), processed by the {\it Gaia} Data Processing and Analysis Consortium (DPAC, \url{http://www.cosmos.esa.int/web/gaia/dpac/consortium}). Funding for the DPAC has been provided by national institutions, in particular the institutions participating in the {\it Gaia} Multilateral Agreement. EM thanks the University of Exeter for support through a Ph.D. studentship. MB acknowledges support from DFG project Kr 2164/15-1. GMK is supported by the Royal Society as a Royal Society University Research Fellow.

\end{document}